\documentclass[amsmath,epsf,eclepsf,epsfig,subfigure,preprint,aps,graphicx,amssymb]{revtex4}
\usepackage[dvips]{graphicx}
\usepackage{color}
\begin{document}
\preprint{HUPD0507}
\title{\Large\bf Low scale Seesaw model and \\
 Lepton Flavor Violating Rare $B$ Decays
 \bigskip\bigskip}
\author{
T. Fujihara$^{(a)}$
 \footnote{fujihara@theo.phys.sci.hiroshima-u.ac.jp},
S. K. Kang$^{(b)}$
 \footnote{skkang@phya.snu.ac.kr},
C. S. Kim$^{(a,c)}$
 \footnote{cskim@yonsei.ac.kr},
D. Kimura$^{(a)}$
\footnote{kimura@theo.phys.sci.hiroshima-u.ac.jp}
and
T. Morozumi$^{(a)}$
\footnote{morozumi@hiroshima-u.ac.jp}
\bigskip\bigskip}

\address{
(a): Graduate School of Science, Hiroshima University,
 Higashi-Hiroshima, Japan, 739-8526
\\
(b): Seoul National University, Seoul, Korea
\\
(c): Department of Physics, Yonsei University, Seoul 120-749, Korea
\bigskip\bigskip}

\begin{abstract}
\noindent We study lepton flavor number violating rare $B$ decays,
$ b \to s l_h^{\pm} l_l^{\mp}$, in a seesaw model with low scale
singlet Majorana neutrinos motivated by the resonant leptogenesis scenario.
The branching ratios of inclusive decays $ b \to s l_h^{\pm}
\bar{l_l}^{\mp} $ with two almost degenerate singlet neutrinos at
TeV scale are investigated in detail. We find that there
exists a class of seesaw model in which the branching fractions
of $ b \to s \tau \mu $ and $\tau \to \mu \gamma$ can be as large
as $10^{-10}$ and $10^{-9}$ within the reach of Super B factories, respectively,
without being in conflict with neutrino mixings and mass squared difference of
neutrinos from neutrino data, invisible decay width of $Z$ and the
present limit of $Br(\mu \to e \gamma)$.

\end{abstract}
\maketitle
\newpage
\newcommand{\dis}[1]{\begin{equation}\begin{split}#1\end{split}\end{equation}}
\newcommand{\beqa}[1]{\begin{eqnarray}#1\end{eqnarray}}
\def\sol{\rm sol}
\def\atm{\rm atm}
\def\Im{\rm{Im}}
\def\be{\begin{equation}}
\def\ee{\end{equation}}
\def\bea{\begin{eqnarray}}
\def\eea{\end{eqnarray}}
\def\ben{\begin{enumerate}}
\def\een{\end{enumerate}}
\def\nn{\nonumber}
\def\lsl{ l \hspace{-0.45 em}/}
\def\ksl{ k \hspace{-0.45 em}/}
\def\qsl{ q \hspace{-0.45 em}/}
\def\psl{ p \hspace{-0.45 em}/}
\def\ppsl{ p' \hspace{-0.70 em}/}
\def\dsl{ \partial \hspace{-0.45 em}/}
\def\Dsl{ D \hspace{-0.55 em}/}
\def\matrix{ \left(\begin{array} \end{array} \right) }
\def\ma{m_A}
\def\mf{m_f}
\def\mz{m_Z}
\def\mw{m_W}
\def\ml{m_l}
\def\ms{m_S}
\def\dag{\dagger}
\def\hf{\textstyle{\frac12~}}
\def\hff{\textstyle{\frac13~}}
\def\hfg{\textstyle{\frac23~}}
\def\Arg{{\rm Arg}}
\def\Im{{\rm Im}}
\def\Re{{\rm Re}}
\def\bv{{\bf v}}
\newcommand{\ov}{\overline}
\newcommand{\f}{\frac}
\baselineskip 0.7cm
\bibliographystyle{plain}
\thispagestyle{empty}
\section {Introduction}
Thanks to the worldwide collaborative endeavor for $B$ factories and neutrino experiments,
our understanding
of the flavor mixing phenomena in quark sector as well as in lepton sector has been
dramatically improved over the past few years. Even though
discoveries of neutrino oscillations in solar, atmospheric and
reactor experiments gave a robust evidence for the existence of
non-zero neutrino masses, we do not yet understand mechanisms of how to generate the masses of neutrinos
and why those masses are so small. The most
attractive proposal to explain the smallness of neutrino masses is the
seesaw mechanism \cite{Minkowski,Yana,Gellmann,MohaSen}
in which super-heavy singlet particles are
introduced. One of the virtues of the seesaw mechanism is to
provide us with an elegant way to achieve the observed baryon
asymmetry in our universe via the related leptogenesis \cite{FY}.
However, the typical
seesaw scale is of order $10^{10} \sim 10^{14}$ GeV, which makes it
impossible to probe the seesaw mechanism
at collider experiments in a foreseeable future. Moreover, the leptogenesis at such a high energy scale
meets a serious problem, the so-called  gravitino problem, when it
realizes in  supersymmetric extensions of the standard model. Thus,
it may be quite desirable to achieve the seesaw mechanism as well as
the leptogenesis at a rather low energy scale. In these regards,
scenarios of the resonant leptogenesis, in which singlet neutrinos with
masses of order $1 \sim 10$ TeV are introduced, have been recently proposed \cite{ Pilab, Pilac, Gonzalez, Brancob, Raidal, West, Hambye, Russela}
Interestingly enough, the amplitudes of some flavor violating processes,
which are highly suppressed in usual
seesaw models with super-heavy singlet neutrinos, may be enhanced
with such low scale singlet neutrinos. Thus, it
is worthwhile to examine how we can probe a seesaw model with low scale singlet
neutrinos at collider experiments and to find some experimental evidence for the seesaw model via
probe of lepton flavor violating processes.

In this paper, within the context of a low scale seesaw model based on $SU(2)_L \times U(1)$ \cite{he},
we study quark and lepton flavor violating (QLFV) rare
decay processes such as $ b \to s l_h^{\pm} l_l^{\mp}$  and $l_h \to l_l \gamma$,
where $l_h$ and $l_l$ denote heavy and light charged leptons, respectively.
After the quark flavor changing neutral currents (FCNC) $ b \to s l^+ l^-$
had been discovered  in  $B$ factories \cite{Belle}, it has been naturally expected that
the next generation experiments could probe well
the lepton flavor violating (LFV) processes \cite{Minkowski,blee,inamib,cvetic}
, which may eventually uncover the
mechanism for the generation of small neutrino masses and the leptogenesis as well.
Therefore, the precise predictions for those processes based on well motivated scenarios  are
very useful to find out if such scenarios can describe the nature correctly or not.
Here we focus on the seesaw model
motivated by the resonant leptogenesis scenario with two almost degenerate singlet
neutrinos at TeV scale. In the context of the seesaw mechanism, lowering the singlet mass scale
leads to an undesirable enhancement of the light neutrino masses unless the Dirac neutrino couplings
are guaranteed to be naturally small. However, as shown in \cite{Pila,
PilaUnder},
despite of the low mass scale of singlet Majorana neutrinos we can obtain light neutrino mass spectrum
consistent with the current neutrino data by tuning the phase of the Yukawa-Dirac mass terms
so that the two degenerate singlets contribute to the low energy effective Majorana mass terms
destructively and the lepton number is approximately conserved.
Interestingly enough, in such a scenario  the sizable LFV processes and the
suppression of the lepton number violation required in
the effective Majorana mass for light neutrinos may coexist.
We will show that the amplitudes of quark FCNC and LFV processes can be enhanced
by considering some specific structure of Yukawa-Dirac
and singlet Majorana mass matrices which are required to achieve the resonant leptpogenesis.
We will obtain rather stringent constraints on QLFV processes
by taking the constraints arisen from the invisible decay of $Z$ boson,
neutrino mass-squared differences and lepton flavor mixings measured at neutrino experiments and
the experimental constraints of LFV processes such as $ l_h  \to l_l \gamma$.

The paper is organized as follows:
In section 2, we present the lepton flavor mixings of seesaw models with arbitrary
number of singlet Majorana neutrinos.
The analytical expressions for  the branching fractions of the QLFV rare $B$ decays
are presented and the model independent bound on the branching fractions is obtained.
In section 3, we study the low mass scale scenarios for singlet neutrinos
and build the models in which the quark FCNC and LFV processes are enhanced.
We give the predictions for the QLFV processes by taking into account
the various constraints.  In section 4, we summarize the results.

\section{Lepton flavor mixig of seesaw model and QLFV rare $B$ decay}

The lepton favor mixings of seesaw model are described in detail in Refs. \cite{branco,broncano}.
Here we extend the model to the case with arbitrary number of singlet neutinos
and introduce a convenient decomposition of the Yukawa-Dirac mass term which is useful
to our study.
Let us start with a seesaw model described by following Lagrangian,
\begin{eqnarray}
  {\cal L}_{m}=-y_{\nu}^{ik}\ov{L_i}N_{R_k}\tilde{\phi}
              -y_{l}^{i} \ov{L_i} l_{R_i} {\phi}
-\f{1}{2}\ov{N_{R_k}}^{c} M_{k} N_{R_k}+h.c.,
\label{lagrangian}
\end{eqnarray}
\def\V{{\cal V}}
and the neutrino mass matrix
\bea
M_{\nu} =\left(\begin{array}{cc}\
        0 & m_D \\
        m_D^T & M \end{array} \right),
\eea
where $M$ is a $N \times N$ real
diagonal singlet Majorana neutrino mass matrix and $m_D$ is a Dirac Yukawa mass term.
Then, $(3+N) \times (3+N)$ neutrino mass matrix can be diagonalized by the mixing matrix $\V$
as,
\bea
M^{\rm Diag}_{\nu}=\V^{\dagger} M_{\nu} \V^{\ast}.
\eea
If the seesaw condition, $i.e.$ $ \frac{m_D}{M} < 1$, is satisfied,
$(3+N) \times (3+N)$ unitary matrix $\V$ can be approximately
parameterized as,
\bea
\V=\left(\begin{array}{cc}
          V & {m_D}\frac{1}{M} \\
        -\frac{1}{M} m_D^{\dagger} V & \left(1_N-
        \frac{1}{M}{m_D}^{\dagger} m_D \frac{1}{M} \right)^{\frac{1}{2}} \end{array}\right),
\eea
where $\V$ satisfies unitarity to the order of $\frac{m_D^2}{M^2}$, and
$1_N$ denotes an $ N \times N$ unit matrix.
Here $ 3 \times 3$  submatrix $V$  is not exactly unitary and
can be written  \cite{broncano} as
\bea
V=\left(1-m_D \frac{1}{M^2} m_D^{\dagger} \right)^{\frac{1}{2}}V_0,
\eea
where
$V_0$ is a unitary matrix which diagonalizes the effective Majorana
mass term,
\bea
m_{eff}&=& -m_D  \frac{1}{M} m_D^T, \nn \\
V_0^{\dagger} m_{eff} V_0^{\ast}&=& \mbox{Diag}[n_1,n_2,n_3],
\eea
where $n_i$ are the masses of three light neutrinos.

For convenience, we introduce the following parametrization for $m_D$
\cite{Fujihara},
\bea
m_D=({\bf m_{D1}, m_{D2},...m_{DN}})
&=&\left(\begin{array}{cccc}
{\bf u_1} & {\bf u_2} & .. & {\bf u_N} \end{array}\right)
\left( \begin{array}{cccc} m_{D1}& 0 & 0 &0 \\
                           0 & m_{D2} & 0 &0 \\
                           0 & 0 & .. &  0 \\
                           0 & 0 & 0  &m_{DN}
\end{array} \right), \nn \\
\label{mD}
\eea
where $N$ unit vectors are introduced as,
\bea
{\bf u_I}=\frac{\bf m_{D I}}{m_{D I}}~,
\label{unit}
\eea
with $m_{D I}=|{\bf m_{D I}}|$. By introducing the parameters
of mass dimension $X_I=\frac{m_{DI}^2}{M_I} \ (I=1 \sim N)$, $m_{eff}$
can be written as
\bea
m_{eff}&=&-\sum_{i=1}^N {\bf u_i} X_i {\bf u_i^T}. \label{meff}
\eea
The charged current is associated with $ 3 \times (3+N)$ submatrix of $\V$.
The deviation from the unitarity of $ 3 \times 3 $ matrix $V$ is given as
\bea
\sum_{a=1}^3 V_{ia} V_{j a}^{\ast}&=&\delta_{ij} -\sum_{I=1}^N
\frac{X_I}{M_I} {{u}_{iI} {u}_{jI}^{\ast}}, \nn \\
\sum_{i={\rm e, \mu, \tau}} V_{ia} V_{ib}^{\ast}&=& \delta_{a b}-
\sum_{I=1}^N ( V_0^\dagger U)_{bI} \frac{X_I}{M_I} (U^{\dagger} V_0)_{I a},
\eea
where $U=({\bf u_1},{\bf u_2}, .. ,{\bf u_N})$.

Now we study the QLFV rare $B$ decay processes
$ b \to s l_h^{\pm} l_l^{\mp}$
in the context of the seesaw model we consider.
We denote $l_h$ as a heavy lepton and $l_l$ as a light lepton, and
the possible combinations for  $(l_h^{\pm}, l_l^{\mp})$ are
$(\tau^{\pm},\mu^{\mp}),(\mu^{\pm}, e^{\mp})$ and $
(\tau^{\pm},e^{\mp})$.
By separating the contributions from the light neutrinos
with masses $n_a~ (a=1 \sim 3)$ and ones from the heavy neutrinos
with masses $M_{A-3}~ (A=4,5,...3+N)$,
we can express the amplitude for $ b\to s l_h^- l_l^+$ as
\bea
&&T=\frac{\sqrt{2} G_F \alpha_{QED} }{ 4 \pi s_W^2} \times \nn \\
&& \left(\overline{u_s} \gamma_{\mu}L u_b \right)
\left( \overline{u_h} \gamma^{\mu} L v_l \right)
\sum_{i=u \ c \ t} V_{ib} V_{is}^{\ast}
\left(
\sum_{a=1}^3
V_{h a} V_{l a}^{\ast}
 E(x_i, y_a) + \sum_{A=4}^{(3+N)} {\cal V}_{h A} {\cal V}_{l A}^{\ast}
 E(x_i, y_{A-3})
 \right),\nn \\
\label{Tmat}
\eea
where $u_b$,$u_s$,$u_h$ and $v_l$ denote the spinor of bottom quark,
strange quark, heavy lepton and light anti-lepton
respectively and
$L=\frac{1-\gamma_5}{2}$,
$x_i=\frac{m_i^2}{M_W^2}$, $y_a=\frac{n_a^2}{M_W^2}$,
$y_A=\frac{M_{A}^2}{M_W^2}$.
$E$ is a Inami-Lim \cite{inamia} function presented as,
\bea
E(x,y)&=&-(1+\frac{x y}{4})\frac{f(x)-f(y)}{x-y}
       +(1-\frac{7 x y}{4})
       \frac{x f^{\prime}(x)-y f^{\prime}(y)}{x-y}, \nn \\
{\rm with}~~~~~ f(x)&=&\frac{x \log[x]}{x-1}.
\eea
We may neglect the up quark loop contribution  because of
the smallness of $V_{ub}V_{us}^{\ast}$.
By using the  unitarity relations for leptonic and
quark sectors,
\bea
&& \sum_{a=1}^3 V_{h a} V_{l a}^{\ast}=
-\sum_{A=4}^{3+N} {\cal V}_{h A} {\cal V}_{l A}^{\ast}, \nn \\
&& V_{tb}V_{ts}^{\ast}=-V_{cb}V_{cs}^{\ast},
\eea
we can simplify Eq. (\ref{Tmat}) as follows;
\bea
T= \frac{\sqrt{2} G_F}{ 4 \pi} \frac{\alpha_{QED}}{s_W^2}
\left(\overline{u_s} \gamma_{\mu}L u_b \right)
\left( \overline{u_h} \gamma^{\mu} L v_l \right)
 V_{tb} V_{ts}^{\ast}
\sum_{A=4}^{3+N} {\cal V}_{h A} {\cal V}_{l A}^{\ast}
\left(\bar{E}(x_t,x_c,y_{A-3}) \right),
\label{T}
\eea
 In Eq.(\ref{T}),
we only keep the contributions of the top quark, charm quark
and heavy neutrinos, and
\bea
 && \bar{E}(x_t,x_c,y_{A-3})
 =E(x_t, y_{A-3}) -E(x_t,0)-E(x_c, y_{A-3})+E(x_c,0) \nn \\
 &&\simeq x_t y_{A-3} \left( \frac{3}{4 (1-x_t)(1-y_{A-3})}-
 \frac{(x_t^2-8x_t+4) \log(x_t) }{4(x_t-1)^2 (x_t-y_{A-3})}
 - \frac{(y_{A-3}^2-8y_{A-3}+4)\log(y_{A-3}
 )}{4(y_{A-3}-1)^2(y_{A-3}-x_t)} \right), \nn \\
\label{inamilimf}
\eea
which agrees with the results in Refs. \cite{he,pal}.
The branching fraction of $ b \to s l_h^- l_l^+$ can
be easily obtained as
\bea
{\rm Br}( b \to s {l_h}^- {l_l}^+)&=&Br( b \to c \bar{\nu_e} e)
\frac{|V_{tb} V_{ts}|^2}{|V_{cb}|^2}
 \frac{P(\frac{m_h}{m_b})}{P(\frac{m_c}{m_b})}
 \left(\frac{\alpha_{QED}}{8 \pi s_W^2} \right)^2
 |\sum_{A=4}^{3+N} {\cal V}_{h A} {\cal V}_{l A}^{\ast} \bar{E}(x_t,x_c,y_{A-3})|^2,\nn \\
\label{branch}
\eea
where the phase factor $P$ is given by
$P(x)=1-8 x^2 + 8 x^6 -x^8-24 x^4 \log(x) $
and $m_h$ denotes the heavier lepton ($l_h$) mass.
For numerical computation, we take
${\rm Br}(b \to c \bar{\nu_e} e) = 0.107, \
 m_b=4.75~(\rm GeV), \
m_c=1.25~(\rm GeV), \
\alpha_{QED}=\frac{1}{137}, \
m_{\tau}=1.78~(\rm GeV),\
m_{\mu}=0.106~(\rm GeV),\
m_{W}=80.4~(\rm GeV).
$
The branching fractions are then given as
\bea
{\rm Br}( b \to s {\tau}^- {e}^+)&=&  1.0 \times 10^{-7} |S(\tau,e)|^2, \nn \\
{\rm Br}( b \to s {\tau}^- {\mu}^+)&=&  1.0 \times 10^{-7} |S(\tau,\mu)|^2, \nn \\
{\rm Br}( b \to s {\mu}^- e^+)&=&
2.8 \times 10^{-7} |S(\mu,e)|^2,
\label{pre}
\eea
where
$S(h,l)$ is the suppression factor defined by
\bea
 S(h,l)&=&\sum_{A=4}^{3+N} {\cal V}_{h A} {\cal V}_{l A}^{\ast}
 \bar{E}(x_t,x_c,y_{A-3}), \nn \\
 &=& \sum_{I=1}^N \frac{X_I}{M_I} {\bf u}_{h I} {\bf u^{\ast}}_{l I}
 \bar{E}(x_t,x_c, y_I).
\label{rel2}
\eea

As can be seen above, the QLFV processes depend on $\frac{X_I}{M_I}=\frac{m_{DI}^2}{M_I^2},\
{\bf u_{hI} u_{lI}^{\ast}} $ and $M_I$ in $\bar{E}$.
If $\frac{X_I}{M_I}$ is not very
small, there will be a chance to detect the QLFV processes
at $B$ factories near future.
First, we can roughly estimate  the
branching fractions of the QLFV processes.
We notice that a model independent constraint on
$\sum_{I=1}^N \frac{X_I}{M_I}$
can be obtained from invisible decay width of $Z$ and charged
curents lepton universality test \cite{loinaz,gla,loinazb}
because
$Z \to \nu_a \bar{\nu_b}$ coupling is suppressed
compared with standard model prediction as,
\bea
\frac{g}{2 \cos \theta_W} \sum_{a,b=1}^3 Z_{ba} Z^{\mu} \bar{\nu_b} \gamma_{\mu}\frac{(1-\gamma_5)}{2}{\nu_a},
\eea
where $Z_{ba}$ is related to the violation of the unitarity of $V$,
\bea
Z_{ba}=\sum_{i= e \mu \tau} V_{ib}^{\ast}V_{ia}=
\delta_{ab}-(V_0^{\dagger}U)_{bI}\frac{X_I}{M_I} (U^{\dagger} V_0)_{Ia}.
\label{ZFCNC}
\eea
In the model under consideration, the effective
number of light neutrinos $N_{\nu}$ is given by
\bea
N_{\nu}=
\sum_{a,b=1}^3
|Z_{ b a}|^2=3-2 \sum_{I=1}^N \frac{X_I}{M_I}~.
\eea
 {}From the experimental result \cite{LEP}, $N_{\nu}=2.984\pm 0.008$,
we can obtain the following bound, \bea \sum_{I=1}^N
\frac{X_I}{M_I} =0.008 \pm 0.004. \label{XM} \eea If the bound is
dominated by N degenerate singlet heavy neutrinos with $X_1 \sim
X_2 \sim X_{N}$, $\frac{m_{DI}}{M_I} \simeq \frac{0.1}{\sqrt{N}}$,
which is achieved in the case of low scale  singlet neutrinos.

\begin{figure}
\includegraphics{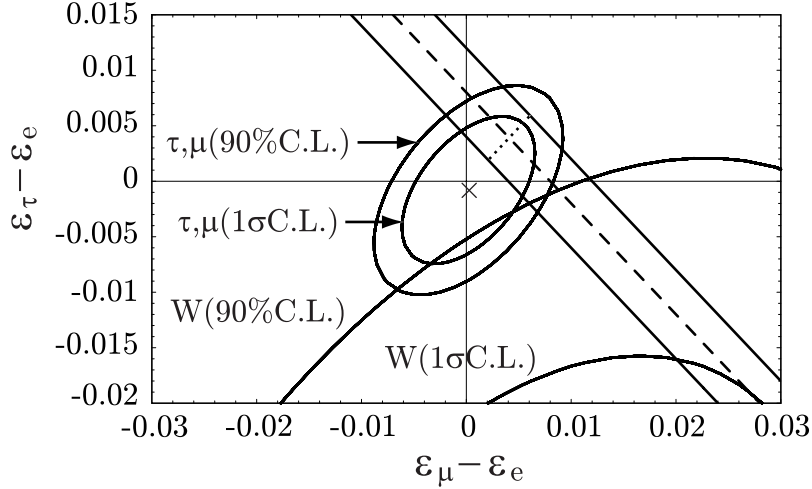}
\caption{The experimentally allowed region of
 $(\epsilon_{\mu}-\epsilon_e, \epsilon_{\tau}-\epsilon_e)$
is shown. $\tau,\mu$ denote the constraints obtained from
$ \tau $ $\mu$ leptonic decays. $W$ denotes the bound
obtained from $ W \to l_i \bar{\nu}$ decays.
The invisible decay width constraints Eq.(\ref{zepsilon}) is also
shown for  $\epsilon_e=0$. The dotted line coresponds to the
prediction of
class B model (NH) case. (See section IV in text.) }
  \label{corr}
\end{figure}

From the fact that the lepton flavor universality of charged
current is generally violated in seesaw model, one can obtain experimental
bounds on $\epsilon_i \equiv \sum_{I=1}^N |u_{iI}|^2 \frac{X_I}{M_I}$
($i=e, \mu ,\tau$) from the lepton universality test of charged
current interactions.
The deviation from the universality is expressed in terms
of the flavor dependent coupling $g_i$ defined as
\bea
\frac{g_i^2}{g^2}&=&\sum_{a=1}^3
|V_{ia}|^2=1-\epsilon_i.
\eea
Using the definition of $g_i$ above, for instance,
the decay width of W into charged lepton $l_i$ and neutrino
is given by,
\bea
\sum_{a=1}^3 \Gamma[ W \to l_i \bar{\nu_a}]&=&\frac{g^2 M_W}{48 \pi}
\sum_{a=1}^3 |V_{ia}|^2 (1-\frac{m_i^2}{M_W^2})^2(1+\frac{m_i^2}{2 M_W^2}),
 \nn \\
&=&\frac{g_i^2 M_W }{48 \pi}
(1-\frac{m_i^2}{M_W^2})^2(1+\frac{m_i^2}{2 M_W^2}),
\eea
where we ignore the neutrino masses.
The experimental bounds on $\epsilon_{\mu}-\epsilon_{e}$
and $\epsilon_{\tau}-\epsilon_{e}$ was obtained
from the ratios of the branching fractions of W decays and also from
the ratios of the branching fractions
of $\tau \to \mu(e) \nu \bar{\nu}$  and $ \mu \to e \nu \bar{\nu}$ decays.
The constraints obtained from $\tau$ and $\mu$ leptonic decays (W leptonic
decays)
are summarized in Eq.(13) (Eq.(6)) of \cite{loinazb},
\bea
\epsilon_{\mu}-\epsilon_{e}&=& 0.0002 \pm 0.0042 \ ( 0.002 \pm 0.022),\nn \\
\epsilon_{\tau}-\epsilon_{e}&=& -0.0008 \pm 0.0044 \ (-0.058 \pm 0.028).
\label{epsilons}
\eea
The off-diagonal elements of the correlation matrix of
$\epsilon_{\mu}-\epsilon_{e}$ and $\epsilon_{\tau}-\epsilon_e$
are 0.51 and 0.44, respectively \cite{loinazb}.
In Fig.(\ref{corr}), we show the constrains Eq.(\ref{epsilons})
on the plane
$(\epsilon_{\mu}-\epsilon_e, \epsilon_{\tau}-\epsilon_e)$
by taking into account the correlations.
In the figure, we also show the constraint obtained from
Eq.(\ref{XM}) for the case that $\epsilon_e $ is vanishing.
The constraint of Eq.(\ref{XM}) is written with $\epsilon_i$ as
\bea
(\epsilon_{\mu}-\epsilon_{e}) + (\epsilon_{\tau}-\epsilon_e)=
(0.008\pm 0.004)-3 \epsilon_e.
\label{zepsilon}
\eea
From Fig.\ref{corr}, we can see that the constraints obtained from
$\tau$ and $\mu$ leptonic decays are consistent with Z invisible decay width
constraint within $1 \sigma$ CL under the assumption  $\epsilon_e=0$
while the constraints
obtained from $W$ decays are not consistent and
$\epsilon_{\tau}-\epsilon_e < 0$ seems to be required in this case.
We come back to the lepton universality constraints when we consider
the specific structure of Yukawa-Dirac mass term in the following sections.

\begin{figure}
 \includegraphics{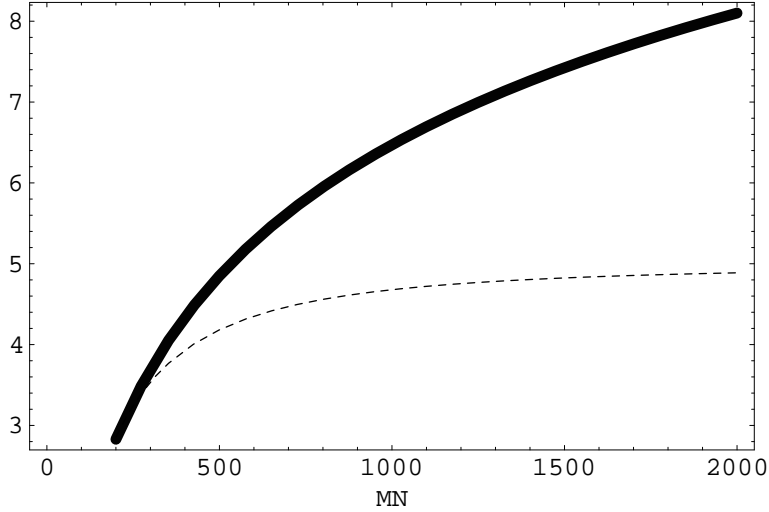}
    \caption{Inami-Lim fiunctions $-\bar{E}$ (thick solid line)
    and $10 \times F_2$ (dashed line) as  functions of the singlet
    neutrino mass $M_N$ (GeV).
     }
  \label{INAMILIM}
  \end{figure}

The Inami-Lim function $\bar{E}(\frac{M_I^2}{m_W^2})$
is shown in Fig. \ref{INAMILIM}.
The typical values for the Inami-Lim function are
$
|{\bar{E}}|\sim 2.8 \sim 8.1  \ {\rm for} \ M=200 \sim 2000~ ({\rm GeV})$.
Finally, from the fact that the factor $u_{h I} u_{l I}^{\ast} $
depends on the flavor structure of $m_D$, the
following relation
\bea
2 |u_{hI} u_{lI}| \le |u_{hI}|^2+|u_{lI}|^2 \le
\sum_{i=e,\mu,\tau}|u_{iI}|^2=1,
\eea
leads us to the constraint,
$|u_{hI} u_{lI}|^2 \le 0.25$. Then, numerical value of $S(h,l)$
for the case with $N$ degenerate $M_i$ and $X_i$ is
\bea
|S(h,l)|^2
 \le \left(N  \frac{X}{M} \right)^2 \bar{E}\left(\frac{M^2}{M_W^2}
 \right)^2 0.25
\simeq 2 \times  10^{-4} \ \sim  2
\times 10^{-3},
\eea
where we denote $M_i \equiv M (i=1 \sim N)$ and $X_i \sim X (i=1 \sim N)$.
The upper bound of the branching
fractions for $ b \to s l_h^{\mp} l_l^{\pm}$ is roughly predicted
to be $10^{-11} \sim 10^{-10}$ for $ 200~({\rm  GeV})< M <
2000~({\rm GeV})$.

\section{Leptonic FCNC and QLFV rare decays with
low mass scale singlet Majorana neutrinos}

We now predict the branching fractions of the QLFV processes
more concretely.
As previously discussed, the branching fractions can be enhanced for
rather large values of $X_I/M_I$.  The large values of  $X_I/M_I$
are realized for low scale of $M_I$, which may not generally be
consistent with neutrino data.
The large values of $X_I/M_I$ can
be consistent with neutrino data when the contributions of $X_I$
to $m_{eff}$ in Eq. (\ref{meff}) are cancelled.
Such a cancellation
can be achieved by taking two almost degenerate small $M_{1,2}$ and
tuning the relative phase between ${\bf u_1}$ and ${\bf u_2}$
so that those two terms contribute to $m_{eff}$ destructively while
keeping $X_3$ so small that its contribution to $m_{eff}$ is suppressed.
Thus, we need some specific flavor structure of Yukawa-Dirac mass term
in order to obtain an enhancement of the branching fractions.
Let us assume $X_1 \sim X_2 \gg X_3$ so as for
$X_{12}\equiv X_1-X_2$ and $X_3$ to be of
order the light neutrino mass squared differences
$\sqrt{\Delta m_{\rm sol}^2}$  or $\sqrt{\Delta  m_{\rm atm}^2}$.
The relative phase of Yukawa-Dirac mass term from
two singlet neutrinos $N_1$ and $N_2$ is tuned as
 ${\bf u_2}=i {\bf u_1}$. Then, $m_{eff}$ becomes
\bea
m_{eff}=-\left({\bf u_1} {\bf u_1^T} X_{12} +{\bf u_3} {\bf u_3^T} X_3 \right).
\label{eff}
\eea
We further assume the orthogonality of ${\bf u_1}$
and ${\bf u_3}$, $i.e.$ ${\bf u_1^{\dagger} \cdot u_3}=0$, so that
${\bf u_1}$ and ${\bf u_3}$ can be directly related to MNS matrix.
Then, there exists a massless state due to the alignment of ${\bf u_1} $ and ${\bf u_2}$,
which is assigned to $n_1=0$ for normal hierarchy
and $n_3=0$ for inverted hierarchy. The other two masses are given by
$X_3$ or $|X_{12}|$.

In Table \ref{class}, we classify the assignment of mass spectrum $(n_1,n_2,n_3)$
and the unitary part of the mixing matrix $V_0$,
\begin{table}
\begin{tabular}{|c|c|c|c|} \hline
 & $(n_1,n_2,n_3)$  & $V_0=({\bf v_1,v_2,v_3}) $ &
  flavor dependence of $ b \to s l_h^- l_l^+$  \\ \hline
Normal & $( 0 \sqrt{\Delta m_{\sol}^2}, \sqrt{\Delta m_{\atm}^2+
\Delta m_{\sol}^2} )$ & &  \\ \hline
Class A & $(0,|X_{12}|, X_3)$ &
       $\left({\bf  u_1^{\ast} \times u_3^{\ast}},{\bf u_1},{\bf u_3}\right)
       p $ & ${\bf v_{h2} v_{l2}^{\ast}}$ \\
Class B & $(0, X_3, |X_{12}|)$ &
       $\left({\bf u_1^{\ast} \times u_3^{\ast}},{\bf u_3}, {\bf u_1}
       \right) p$ & ${\bf v_{h3} v_{l3}^{\ast}}$ \\ \hline
Inverted & $(\sqrt{\Delta m_{\atm}^2-\Delta m_{\sol}^2}, \sqrt{\Delta m_{\atm}^2},0)$ &  & \\ \hline
Class A & $(|X_{12}|, X_3,0)$  & $
\left({\bf u_1},{\bf u_3},{\bf  u_1^{\ast} \times u_3^{\ast}}\right) p $
& ${\bf v_{h1} v_{l1}^{\ast}}$\\\
Class B & $(X_3,|X_{12}|, 0) $ &
$\left({\bf u_3},{\bf u_1}, {\bf u_1^{\ast} \times u_3^{\ast}}\right) p $ &
${\bf v_{h2} v_{l2}^{\ast}}$ \\ \hline
\end{tabular}
\caption{The assignment of mass spectrum and
MNS matrix.
}
\label{class}
\end{table}
where $p$ is a diagonal Majorana phase which is irrelevant
to the QFLV and LFV processes and thus we omit it from now on.
Notice that $V_0$ is identical to the MNS matrix if we neglect
its deviation from $V_{\rm MNS}$, $V_0-V_{\rm MNS}={\cal O}(\frac{m_D^2}{M^2})$.
In fact, since the QLFV and LFV processes are already in the order of
$\frac{{m_D}^2}{M^2}$ at the leading order, the difference can be safely ignored.
The flavor dependence of the amplitudes for the QFLV and LFV processes is then
extracted in terms of the mixing angles of
the neutrino oscillation from $V_0=V_{\rm MNS}$:
\bea
V_0
&&=\left(\begin{array}{ccc}
c_{13} c_{\sol} & c_{13} s_{\sol} & 0 \\
-s_{\sol} c_{\atm}&  c_{\atm} c_{\sol} & c_{13} s_{\atm} \\
s_{\sol} s_{\atm}  &  -s_{\atm} c_{\sol} & c_{13} c_{\atm} \end{array} \right)
                +s_{13} \left(\begin{array}{ccc}
0 & 0 & \exp(-i\delta) \\
-c_{\sol} s_{\atm} \exp(i \delta) & -s_{\sol} s_{\atm} \exp(i \delta) & 0 \\
-c_{\sol} c_{\atm} \exp(i \delta) & -s_{\sol} c_{\atm} \exp(i \delta)&
0  \end{array}
 \right), \nn \\
\eea
where we take $s_{\sol}=0.56,~~ c_{\sol}=0.84 $ and $s_{\atm}=c_{\atm}=\sqrt{\frac{1}{2}}$.

In Table \ref{flavor},
we present the relevant combinations of $\bv_{ij}$ which correspond
to the flavor dependence shown in the fourth
column of Table I. The value of $s_{13}$ is very small and thus we ignore
the terms of order ${\cal O}(s^2_{13})$.
\begin{table}
\begin{tabular}{|c|c|c|} \hline
$ {\bv}_{\mu 2} \bv_{e2}^{\ast}$ & $ c_{13} s_{\sol} c_{\sol} c_{\atm}$ & $0.33c_{13}$\\
$ \bv_{\tau 2} \bv_{\mu 2}^{\ast}$ &$-c_{\atm} s_{\atm} c_{\sol}^2$ & $-0.35$\\
$\bv_{\tau2} \bv_{e 2}$&$-c_{13} s_{\atm} c_{\sol} s_{\sol}$ & $-0.33 c_{13}$\\
\hline
$\bv_{\mu 3} \bv_{e 3}^{\ast}$ &$ c_{13} s_{\atm} s_{13} \exp(i \delta) $&
$0.71 s_{13}c_{13} \exp(i \delta)$\\
$\bv_{\tau 3} \bv_{\mu 3}^{\ast}$ &$c_{13}^2 s_{\atm} c_{\atm} $ &
$0.5 c_{13}^2$\\
$\bv_{\tau 3} \bv_{e 3} $&$ c_{13} c_{\atm} s_{13} \exp(i \delta)$ &
$0.71 s_{13}c_{13} \exp(i\delta)$ \\ \hline
$\bv_{\mu 1} \bv_{e 1}^{\ast}$ & $-c_{13} c_{\sol} s_{\sol} c_{\atm}$ & $
-0.33 c_{13}$\\
$\bv_{\tau 1} \bv_{\mu 1}^{\ast}$&$ -s_{\sol}^2 c_{\atm} s_{\atm} $ &$-0.16$ \\
$\bv_{\tau 1} \bv_{e 1}^{\ast}$& $c_{13} c_{\sol} s_{\sol} s_{\atm}$ & $0.33
c_{13}$  \\
\hline
\end{tabular}
\caption{The combinations of $\bv_{ij}$ relevant to the flavor dependence
of QLFV and LFV decays.}
\label{flavor}
\end{table}
Then, the suppression factor $S(h,l)$ is approximately given by
\bea
S(h,l)
&\simeq&\left(\frac{X}{M_1}
\bar{E}(x_t,x_c,y_1) +\frac{X}{M_2} \bar{E}(x_t,x_c,y_2) \right)
{\bf v_{h \alpha}} {\bf v_{l \alpha}^{\ast}}, \label{spfac}
\eea
where $\alpha$ denotes the index depending on
the class and neutrino mass hierarchy, $X_1 \simeq X_2 \equiv X$. And
the term proportional to
$\frac{X_3}{M_3}$
is not relevant at all and thus ignored.
By using Eq. (\ref{spfac}), the ratios of the branching fractions are given by
\bea
&&\frac{{\rm Br}(b \to s \tau^{\pm} e^{\mp})}{{\rm Br}( b \to s \tau^{\pm} \mu^{\mp})}
=\left|\frac{\bv_{\tau \alpha}\bv^{\ast}_{e\alpha}}{\bv_{\tau \alpha}
\bv^{\ast}_{\mu\alpha}}\right|^2, \label{br1}\\
&&\frac{{\rm Br}(b \to s \mu^{\pm} e^{\mp})}{{\rm Br}( b \to s \tau^{\pm} \mu^{\mp})}
=\frac{P(\frac{m_\mu}{m_b})}{P(\frac{m_\tau}{m_b})}
\left|\frac{\bv_{\mu \alpha}\bv^{\ast}_{e\alpha}}
{\bv_{\tau \alpha}\bv^{\ast}_{\mu\alpha}} \right|^2,  \label{br2}
\eea
where $P(\frac{m_\mu}{m_b})$, and $P(\frac{m_\tau}{m_b})$ are phase space factors and
$\frac{P(\frac{m_{\mu}}{m_b})}{P(\frac{m_{\tau}}{m_b})}=2.74$.

In Table III, the numerical results of the ratios of the branching
fractions given in Eqs. (\ref{br1},\ref{br2}) are presented.
\begin{table}
\begin{tabular}{|c|c|c|c|} \hline
 & Class A NH (Class B IH) & Class B NH  & Class A IH  \\ \hline
$\frac{{\rm Br}(b \to s \tau^{\pm} e^{\mp})}{{\rm Br}( b \to s \tau^{\pm} \mu^{\mp})}$ &
$\left(\frac{s_{\atm}c_{\sol}s_{\sol}}{c_{\atm}s_{\atm} c_{\sol}^2}\right)^2= 0.89 $ &
$\left(\frac{c_{\atm}s_{13}}{s_{\atm} c_{\atm}} \right)^2=
2.0s^2_{13}$ & $\left(\frac{c_{\sol} s_{\sol} s_{\atm}}
{s_{\sol}^2 c_{\atm} s_{\atm}} \right)^2=4.5$ \\ \hline
$\frac{{\rm Br}(b \to s \mu^{\pm} e^{\mp})}{{\rm Br}( b \to s \tau^{\pm} \mu^{\mp})}$&
 $ \left(\frac{s_{\sol} c_{\sol} c_{\atm}}
{c_{\atm}s_{\atm} c_{\sol}^2}\right)^2
 \frac{P_{\mu}}{P_{\tau}}=2.4$ & $
    \left( \frac{s_{\atm} s_{13}}{s_{\atm}c_{\atm}} \right)^2
\frac{P_{\mu}}{P_{\tau}}=5.5s^2_{13}$ & $
\left(\frac{c_{\sol} s_{\sol} c_{\atm}}{ s_{\sol}^2 c_{\atm} s_{\atm}}
\right)^2 \frac{P_{\mu}}{P_\tau}=12.0 $ \\ \hline
\end{tabular}
\caption{Ratios of the branching fractions of $ b \to s l_h^- l_l^+$.
$P_{\mu}\equiv P(\frac{m_{\mu}}{m_b})$ and $P_{\tau} \equiv
P(\frac{m_{\tau}}{m_b})$.
}
\label{ratbr}
\end{table}
\begin{table}
\begin{tabular}{|c|c|c|c|} \hline
& Class A NH (Class B IH) & Class B NH & Class A IH \\ \hline
$\frac{{\rm Br}(\tau\to e \gamma)}{{\rm Br}(\tau \to \mu \gamma)}$ &
$\left(\frac{s_{\atm}c_{\sol}s_{\sol}}{c_{\atm}s_{\atm} c_{\sol}^2}\right)^2=0.89$ &
$ \left(\frac{c_{\atm}s_{13}}{s_{\atm} c_{\atm}} \right)^2
 = 2.0 s_{13}^2$ & $
\left(\frac{c_{\sol} s_{\sol} s_{\atm}}
{s_{\sol}^2 c_{\atm} s_{\atm}} \right)^2 =4.5$\\ \hline
$\frac{{\rm Br}(\mu \to e \gamma)}{{\rm Br}(\tau \to \mu \gamma)}  $
&
$\left(\frac{s_{\sol} c_{\sol} c_{\atm}}
{c_{\atm}s_{\atm} c_{sol}^2}\right)^2 \frac{\tau_{\mu}}{\tau_\tau}
 \left(\frac{m_{\mu}}{m_{\tau}}\right)^5=5.0$ &
$ \left( \frac{s_{\atm} s_{13}}{s_{\atm}c_{\atm}} \right)^2
\frac{\tau_{\mu}}{\tau_{\tau}} \left(\frac{m_{\mu}}{m_{\tau}}
\right)^5=11 s_{13}^2 $ &
$ \left(\frac{c_{\sol} s_{\sol} c_{\atm}}{ s_{\sol}^2 c_{\atm} s_{\atm}}
\right)^2 \frac{\tau_{\mu}}{\tau_\tau} \left(\frac{m_{\mu}}{m_{\tau}}
\right)^5=25$ \\ \hline
\end{tabular}
\caption{Ratios of the branching fractions of $ l_h \to l_l \gamma$.}
\label{flavorg}
\end{table}
It can be seen that in Class B model for NH case, only $ b \to s \tau^- \mu^+$ can be
much larger than the other channel because
of the absence of the  suppression factor $s_{13}^2$ for $ \tau \mu$
final states, while
the branching fractions
of the different channels in models except Class B for NH case are within a factor of 10.
Furthermore, as discussed in Ref. \cite{he}, there is strong correlation between
QLFV processes and LFV radiative decays
$l_h \to l_l \gamma $.  Experimentally, there are stringent bounds
as ${\rm Br}( \mu \to e \gamma)< 1.2 \times 10^{-11}$ \cite{MEGA},
${\rm Br}( \tau \to \mu \gamma)< 6.8  \times 10^{-8}$ \cite{Babartaumu} and
${\rm Br}( \tau \to e \gamma)< 1.1 \times 10^{-7} (3.9 \times
10^{-7}) $  \cite{Babartaue,Belletaue}.
The bounds on the LFV processes
stringently constrain the branching fractions for QLFV processes.

We note that the bound on $\frac{X}{M}$ from the
invisible decay width of $Z$ and the present upper bound on
$ \mu \to e \gamma$ are not compatible with each other for Class A and
Class B IH case, as shown in Table IV.
The branching fraction for $ l_h \to l_l \gamma$
is,
\bea
{\rm Br}(l_h \to l_l \gamma)=\frac{\alpha^3}{256 \pi^2 s_W^4} \frac{m_h^4}{M_W^4}
\frac{m_h}{\Gamma_h} |G|^2.
\eea
Numerically computing the pre-factors, the branching fractions are given by
\bea
{\rm Br}(\tau \to \mu \gamma)&=& 5.4 \times 10^{-4} |G(\tau ,\mu)|^2, \nn \\
{\rm Br}(\tau \to e \gamma)&=& 5.4 \times 10^{-4} |G(\tau, e)|^2, \nn \\
{\rm Br}(\mu \to e \gamma)&=& 3.1 \times 10^{-3} |G(\mu, e)|^2,
\label{brg}
\eea
where $G$ is a suppression factor defined by
\bea
G&=&\sum_{\alpha=4}^{3+N} V_{l \alpha} V_{h \alpha}^{\ast} F_2(y_\alpha)
\nn \\
&=& \sum_{I=1}^{N} u_{lI} u_{h I}^{\ast} \frac{X_I}{M_I} F_2(y_I),
\eea
where $y_I=\frac{M_I^2}{m_W^2}$ and $F_2$ is  Inami-Lim function, $
F_2(y)=-\frac{2 y^3 + 5 y^2 -y}{4 (1-y)^3} -\frac{3 y^3 \log y}{2(y-1)^4}$,
as shown in Fig. \ref{INAMILIM}.

With $M_1=M_2=M \ll M_3$ and $X_1 \simeq X_2 =X \gg X_3$,
for Class A and Class B (IH) model, we predict,
\bea
{\rm Br}(\mu \to e \gamma) = 3.1 \times 10^{-3} \left(\frac{2X}{M}\right)^2 (0.33)^2
F_2(y)^2 \ge 4.4 \times 10^{-10},
\label{mueg}
\eea
where we have used $\frac{2 X}{M} \ge 0.004$ and $M \ge 200$ (GeV).
Therefore, Class A and Class B (IH) models are excluded.
If the  bound on $\frac{X}{M}$ in Eq. (\ref{XM}) is not taken into
account, one can obtain
from Eqs. (\ref{pre},\ref{brg}),
\bea
{\rm Br}(b \to s \tau^- \mu^+)&=&
1.9\times  10^{-4} {\rm Br}(\tau \to \mu \gamma)
\frac{|S(\tau \mu)|^2}{|G(\tau \mu)|^2} \le 1.3 \times 10^{-11}
\left|\frac{S(\tau \mu)}{G(\tau \mu)} \right|^2,
 \nn \\
{\rm Br}(b \to s \tau^- e^+)&=&
1.9 \times 10^{-4} {\rm Br}(\tau \to e \gamma) \frac{|S(\tau e)|^2}{|G(\tau e)|^2}
\le 2.0 \times 10^{-11} \left|\frac{S(\tau e)}{G(\tau e)}\right|^2, \nn \\
{\rm Br}(b \to s \mu^- e^+)&=&
9.0 \times 10^{-5} {\rm Br}(\mu \to e \gamma) \frac{|S(\mu e)|^2}{|G(\mu e)|^2} \le
1.1 \times 10^{-15} \left|\frac{S(\mu e)}{G(\mu e)} \right|^2.
\label{brc}
\eea
While Eq. (\ref{brc}) depends neither on the mass spectrum
of heavy Majorana neutrinos nor on the flavor structure of Yukawa-Dirac mass
terms, the ratio of $|\frac{S}{G}|$ depends on the details of them.
For the present case
 with $M_1 =M_2=M \ll M_3 $ and $X_1=X_2=X \gg X_3$, the ratio
 $|S(h,l)/G(h,l)|$ is simply given as,
\bea
\left|\frac{S(h,l)}{G(h,l)}\right|^2=
\left|\frac{\bar{E}(x_t,x_c,y)}{F_2(y)} \right|^2=98~(M=200~{\rm GeV})
\sim 285~ (M=2000~{\rm GeV}).
\eea
Therefore, the upper bounds on the branching fractions given in Eq. (\ref{brc})
are translated to
\bea
{\rm Br}(b \to s \tau^- e^+) && \le (1.3  \sim  3.6) \times 10^{-9} \nn \\
{\rm Br}( b \to s \tau^- \mu^+) && \le (2.0 \sim 5.0) \times 10^{-9} \nn \\
{\rm Br}(b \to s \mu^- e^+) && \le (1.1 \sim 3.0) \times 10^{-13}.
\label{upper}
\eea
The range of the upper bounds corresponds to $M=200\sim 2000$ (GeV).
For Class A and Class B with IH case,  by combining
$\mu \to e \gamma$ upper limit,
we may obtain much tighter
bound on the other LFV and QLFV processes.
Let us consider Class A (NH) and Class B (IH) cases.
From Table \ref{ratbr} and Table \ref{flavorg},  the upper bounds
on  ${\rm Br}( \mu \to e \gamma)$ and
${\rm Br}( b \to s \mu^- e^+)$ obtained in
Eq. (\ref{upper}) severely constrain the other modes,
\bea
{\rm Br}( b \to s \tau^- e^+) && \le (0.41 \sim 1.1)\times 10^{-13}, \nn \\
{\rm Br}( b \to s \tau^- \mu^+) && \le (0.46 \sim 1.25) \times 10^{-13}, \nn \\
{\rm Br}(\tau \to \mu \gamma) && \le 2.4 \times 10^{-12}, \nn \\
{\rm Br}(\tau \to e \gamma) && \le  2.1 \times 10^{-12}.
\label{boundall}
\eea
The bounds similar to Eq. (\ref{boundall}) are obtained for Class A IH case,
\bea
{\rm Br}( b \to s \tau^- e^+) && \le (0.41 \sim 1.1)\times 10^{-13},\nn \\
{\rm Br}( b \to s \tau^- \mu^+) && \le (0.92 \sim 2.5) \times 10^{-14}, \nn \\
{\rm Br}(\tau \to \mu \gamma) && \le  4.8 \times 10^{-13},\nn \\
{\rm Br}(\tau \to e \gamma) && \le  2.2 \times 10^{-12}.
\eea

We note that the upper bound of the branching fractions of QLFV
processes for Class A (NH, IH)
 and Class B (IH) models
are $10^{-14} \sim 10^{-13}$
and the branching fractions for LFV processes is $10^{-13} \sim 10^{-12}$.
As we have already shown in Eq.(\ref{mueg}), 
the Class A model and Class B model
for IH case can not satisfy the upper limit of  the branching fraction
of $\mu \to e \gamma$ and  constraint from
the effective light neutrinos number
$N_{\nu}$ simultaneously. This is  because the former requires
very small $\frac{X}{M}$, while the latter requires larger $\frac{X}{M}$.
Below, we show
Class B model for NH case may satisfy  both constraints.
Furthermore, the model predicts the large branching
fractions of $\tau \to \mu \gamma$ and
$b \to s \tau \mu$ which are within the reach of near future Super $B$ factories
\cite{Superbelle,Superbabar}.
If we take into account the constraints coming from ${\rm Br}(\mu \to e \gamma)$
and the effective number of light neutrinos $N_{\nu}$,
we can have the parameter regions consistent with
the present bounds only for Class B model  with NH case.
In this class, the stringent experimental limit
on
${\rm Br}(\mu \to e \gamma)$ is not effective on $ {\rm Br}(\tau \to \mu \gamma)$
and ${\rm Br}(b \to s \tau \mu)$, because the former process is proportional
to $s_{13}^2$ and thus ignorable, but the latter processes are not suppressed by the
factor.

\begin{figure}
\includegraphics{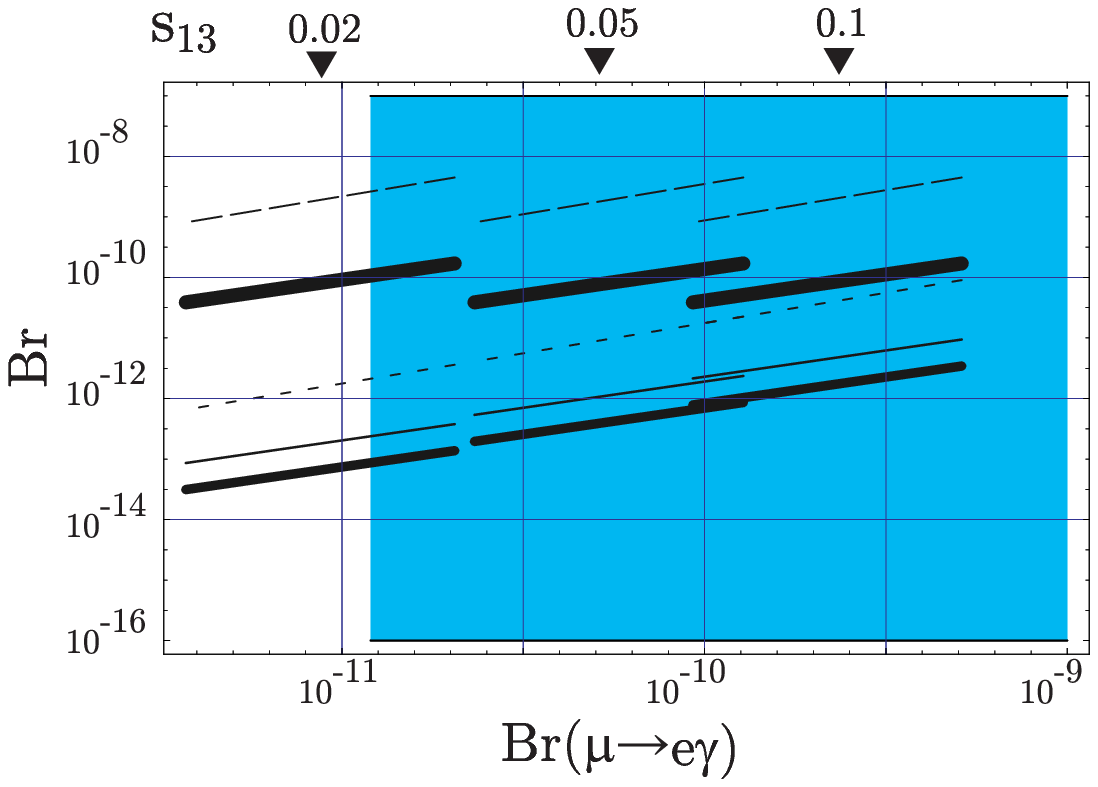}
    \caption{Correlation between the
    branching fraction for $\mu \to e \gamma$ and the branching
     fractions for $ b \to s \tau \mu $ (thick solid line) $  b \to s \tau e$
     (solid line) $  b \to s \mu e$ (thin solid line), $\tau \to \mu \gamma$
     (long dashed line) $ \tau \to e \gamma$ (dashed line) for Class B model with NH case.
    From left to right, the lines correspond to $s_{13}=0.02,0.05,0.1$, repectively.
    The shaded region is excluded by the current bound on $ Br(\mu \to e \gamma)$.
     }
  \label{Class BNH}
\end{figure}

In Fig. \ref{Class BNH}, we have shown the correlation of branching
fractions between $\mu \to e \gamma$
and the other LFV and QLFV processes in  Class B model for NH case.
The numerical results in Fig. \ref{Class BNH} are obtained as follows:
We first set
$X_1 \sim X_2=X$ and $M_2=M_1=M$. From the constraint given in Eq. (\ref{XM}),
the allowed range of $M$ is ,
\bea
\frac{m_{D1}}{\sqrt{\sigma_{\rm max}}} \sqrt{2} <
M < \frac{m_{D1}}{\sqrt{\sigma_{\rm min}}} \sqrt{2},
\label{range}
\eea
with $\sigma_{\rm min}=0.004, \sigma_{\rm max}=0.012$.
When we fix $m_{D1}$,
the allowed
range of $M$ is determined. By varying $M$ within the above
range, we plot the correlation between
${\rm Br}( \mu \to e \gamma)$ and the other five QLFV and LFV
branching fractions. Here, $s_{13}$ is a free parameter
and is chosen to be $0.02,0.05$ and $0.1$. $m_{D1}$ is chosen to be $100$
(GeV). As can be seen in
Fig. \ref{Class BNH}, the present upper limit on
${\rm Br}(\mu \to e \gamma)$ gives a very tight bound on $s_{13}$, typically smaller
than $ 0.02$.
With this small $s_{13}$, $\tau \to e \gamma$ and $b \to s \tau(\mu) e $
are also severely suppressed. Only $ b \to s \tau \mu $ and $\tau \to \mu \gamma$
are free from the suppression and the former branching fraction can
be as large as $10^{-10}$ and the latter can be $ 10^{-9}$. They are
independent on small $s_{13}$.

\begin{figure}
\includegraphics{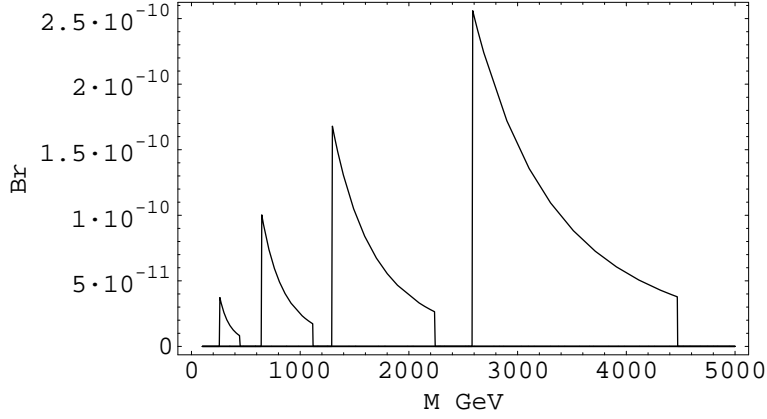}
    \caption {$ {\rm Br}(b \to s \tau \mu)$ {\it vs.}
    $M$ for Class B (NH). From left to right, the curves correspond to
    $m_{D1}=20,50,100,200$ (GeV), respectively.}
\label{MmD}
\end{figure}

\begin{figure}
\includegraphics{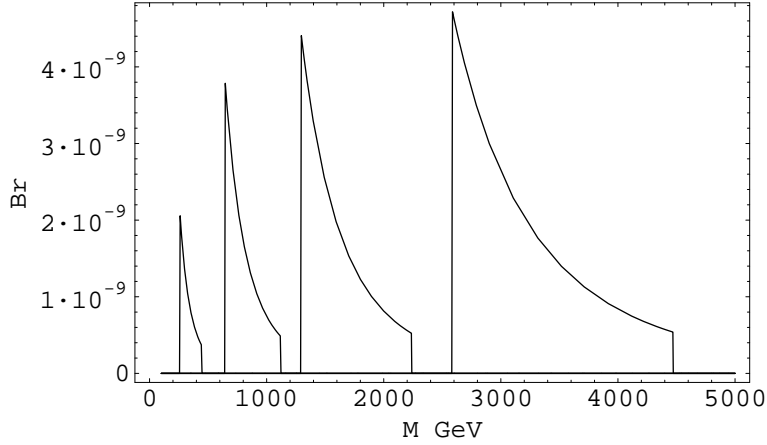}
    \caption{${\rm Br}(\tau \to \mu \gamma)$
    {\it vs.} $M$  for Class B (NH). From left to right,
   the curves correspond to $m_{D1}= 20,50,100,200$ (GeV),respectively.}
  \label{MmDg}
\end{figure}

\begin{figure}
\includegraphics{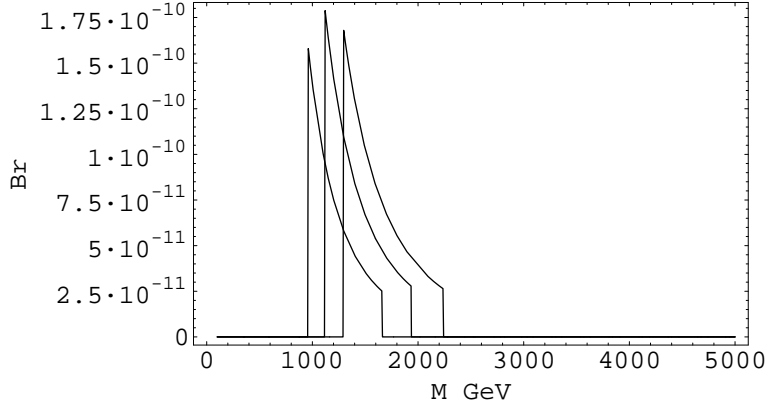}
\caption{${\rm Br}(b \to s \tau  \mu)$ {\it vs.} $M$
     for Class B (NH). We fix $m_{D1}=100$ (GeV).
    From right to left, the curves correspond to the ratio $R=\frac{M_2}{M}$
    $1,2,10$, respectively.}
  \label{RR}
\end{figure}

We show in Fig. \ref{MmD} and Fig. \ref{MmDg} the dependence of
the branching fractions of
$b \to s \tau \mu$ and $\tau \to \mu \gamma$ on $m_{D1}$ and
 the heavy Majorana neutrino mass for
 the exact degenerate case, $i.e.$ $M_1=M_2=M$.
We fix $m_{D1}$ to $20,50,100,200$ (GeV).
Although the branching fractions become small as $m_{D1}$ becomes small,
the change of the branching fractions is within a factor 10.
We also consider the non-degenerate case for Majorana neutrino
masses ($M_1\equiv M \ne M_2$) while keeping the degeneracy  $X_1 \sim X_2=X$.
By setting  $M_2 \equiv  R M_1$ and $M_1 \equiv M$, the dependence of
the branching fraction $ b \to s \tau \mu$
on the ratio $R$ is studied.
The allowed range of the lightest heavy Majorana neutrino mass
$M$ of Eq. (\ref{range})
is modified as $ \frac{m_{D1}}{ \sqrt{\sigma_{\rm max}}}
\sqrt{1+\frac{1}{R}} <M <\frac{m_{D1}}{\sqrt{\sigma_{\rm min}} }
\sqrt{1+\frac{1}{R}}$.
From Fig. \ref{RR}, we find that
 the lower and the upper limits
of $M$ become smaller as $R$ becomes larger. However, the branching fraction
does not change so significantly.
\\
\section{summary and discussion}
As shown, the contributions of the singlet Majorana neutrinos  to
QLFV and LFV decays can be significant in the low scale seesaw
model motivated by resonant leptogenesis.
The branching fractions of inclusive decays $ b \to s l_h^{\pm} l_l^{\mp}$
in the seesaw model considered in this paper depends on
the suppression factor $\frac{m_{D}}{M}$  which is arisen from the mixing between
the singlet heavy neutrinos with three light neutrinos, and
can be as large as about $10 \%$ without being in conflict with the neutrino
mass squared differences from neutrino data and the current bound on the
invisible decay width of $Z$ boson.

We have classified four classes of the model along with the light neutrino
mass spectrum and the assignment of the mixing matrix $V_0$, and studied how
the ratios of the branching fractions for the various channels of QLFV and LFV decays
along with lepton flavors could be distinctively predicted in each class.
We have found that only the class B for NH case presented in Table I survives the current
limit on $Br( \mu \to e \gamma)$ and the invisible decay width of Z boson.
One may check if the model is consistent with the experiments
of lepton universality
tests. The class B for NH case predicts
\bea
\epsilon_e
&=&\frac{2X}{M} s_{13}^2 \le 5 \times 10^{-6} \ (s_{13} \le 0.02), \nn \\
\epsilon_{\mu}&=& \frac{2X}{M} s_{\rm atm}^2 c_{13}^2 \simeq 0.004, \nn \\
\epsilon_{\tau}&=& \frac{2X}{M} c_{\rm atm}^2 c_{13}^2 \simeq 0.004.
\eea
The model predicts very small $\epsilon_e$
and $\epsilon_{\mu}=\epsilon_{\tau}$ which are shown in Fig.(\ref{corr})
with dotted line.
The model is consistent with the constraint of
Z invisible decay width and the lepton universalty
constraints from $\tau$ and $\mu$ decays while
it may not be consistent with the lepton universality constraints
determined by $W$ decays.
In this class,
the branching fractions of $ b \to s \tau \mu$ and
$\tau \to \mu \gamma$ are predicted to be  as large as $10^{-10}$
and $10^{-9}$, respectively. Such large branching fractions can
be tested in the future $B$ factory experiments.
 The enhancment of the branching fractions of QFLV and LFV
 is originated from the one loop Feynman diagrams in which the heavy
 Majorana neutrinos contribute to and
it is non-susy contribution.
\\

\begin{center}
{\bf Acknowledgements}
\end{center}

\noindent We thank K. Homma and K. Ishikawa for useful discussions.
 C.S.K. was supported in part by JSPS,
in part by CHEP-SRC Program,
in part by the Korea Research Foundation Grant funded by the Korean Government (MOEHRD)
No. R02-2003-000-10050-0 and
in part by No. KRF-2005-070-C00030.
T.M. is supported
by the kakenhi of MEXT, Japan, No. 16028213.
S.K.K. was supported in part by BK21 program of the Ministry of
Education in Korea.

\end{document}